\documentstyle[epsf]{aipproc}

\def\MSbar{$\overline{{\rm MS}}$}
\def\gsim{\, \lower0.5ex\hbox{$\stackrel{>}{\sim}$}\, }

\begin{document}
\pagestyle{plain}

\def\figi{
\begin{figure}[t!] 
\centerline{
\epsfxsize=0.98\hsize
\epsfbox{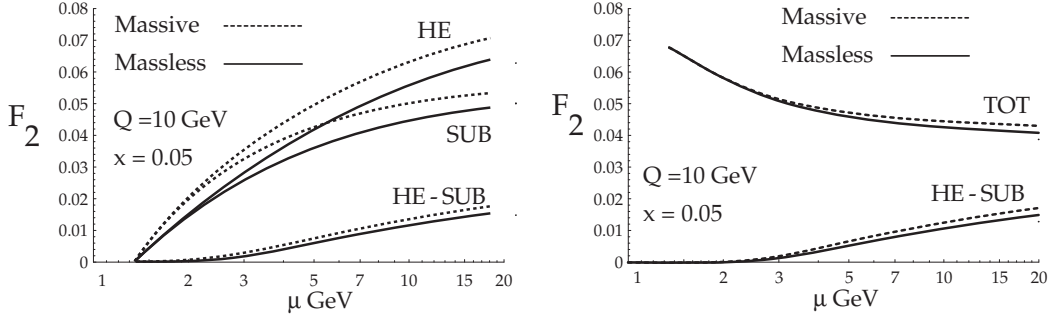}
}
\vspace{-00pt}
\caption{ 
 $F_2$ {\it vs.} $\mu$ for DIS c-production. 
 a) $F_2^{HE}$, $F_2^{SUB}$ and the difference $F_2^{HE}-F_2^{SUB}$.
The  solid curves are for the mass-independent evolution scheme, 
and the   dashed  curves are for the mass-dependent evolution scheme.
 b) $F_2^{TOT}$ and $F_2^{HE}-F_2^{SUB}$. 
 The difference between the mass-independent evolution and 
mass-dependent evolution for $F_2^{TOT}$ is 
higher order and comparable or less than the $\mu$-variation.
}
\vspace{-00pt}
\label{fig1}
\end{figure}
}
\def\figii{
\begin{figure}[t!] 
\centerline{
\epsfxsize=0.49\hsize
\epsfbox{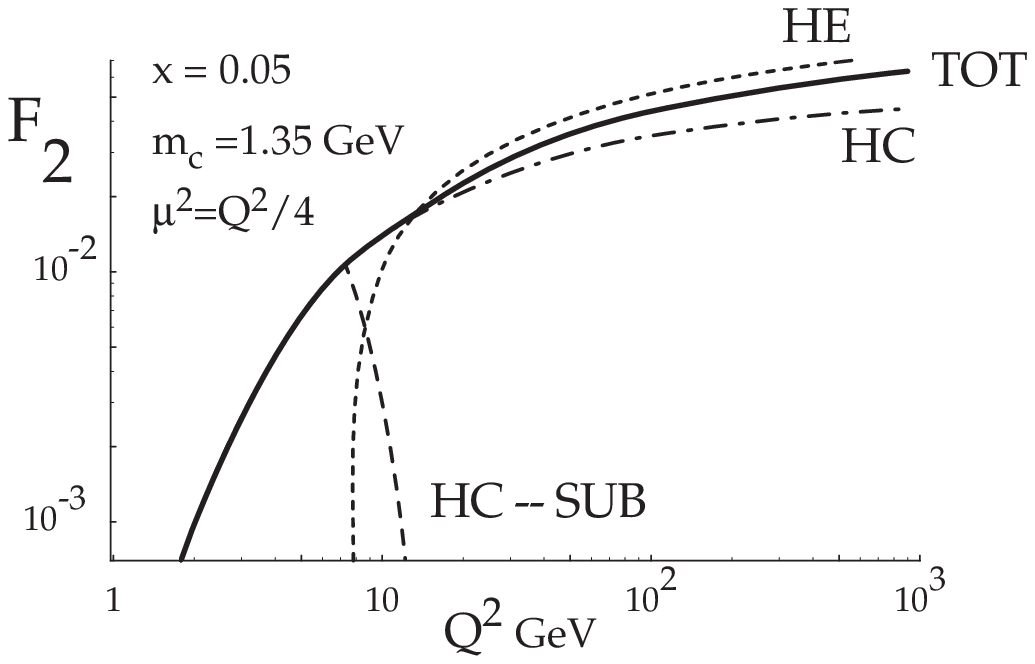}
\hfil
\epsfxsize=0.49\hsize
\epsfbox{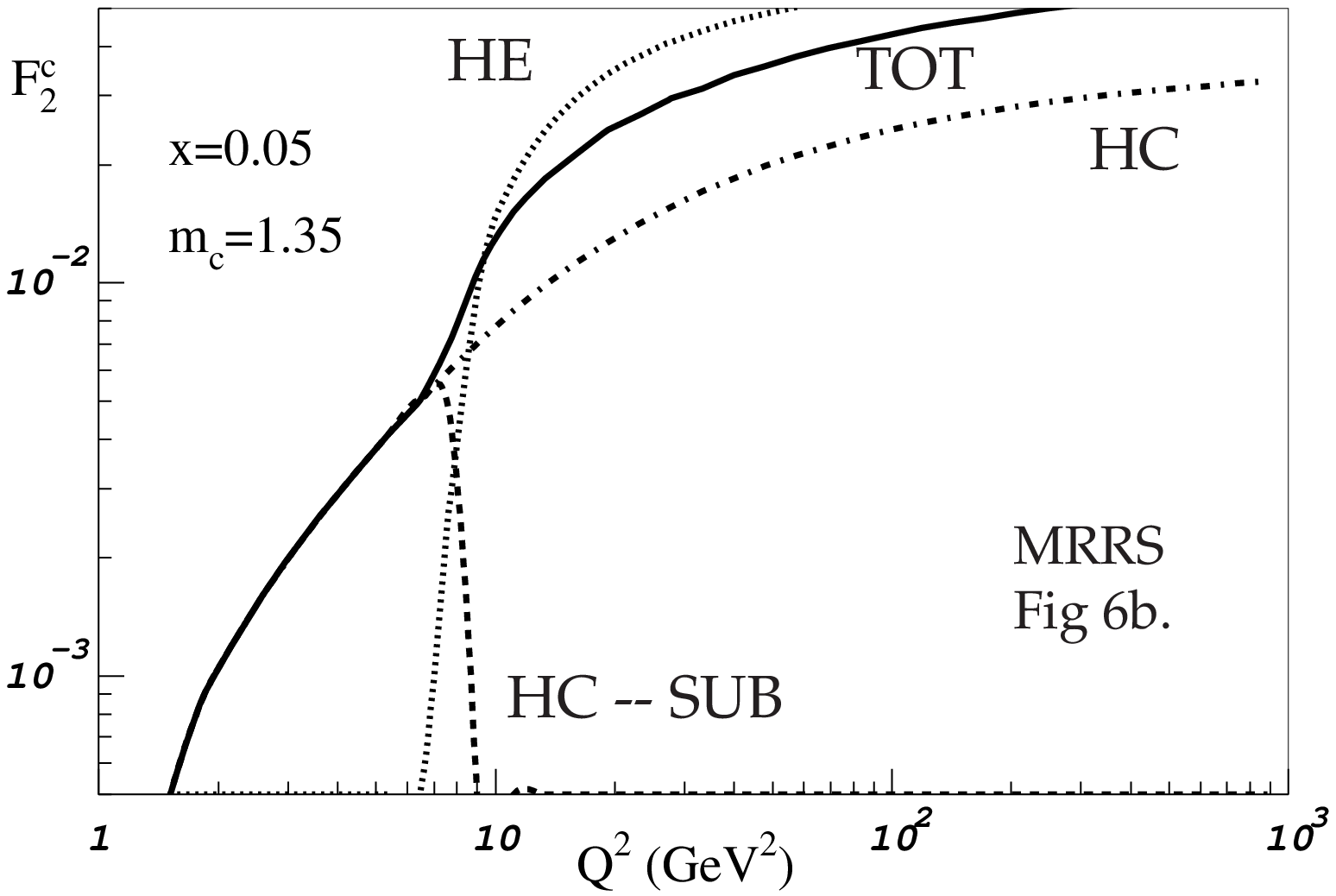}
}
\vspace{-00pt}
\caption{
 $F_2$ {\it vs.} $Q^2$ for DIS
c-production. The scale choice is:  $\mu^2 = Q^2/4$. 
 a) The general mass formalism with mass-dependent evolution.
 b) The MRRS scheme with mass-dependent evolution. 
This figure is taken directly from  
ref.~\protect\cite{MRRS}, Fig.6b.
The curves are re-labeled to correspond to our notation. 
}
\vspace{-00pt}
\label{fig2}
\end{figure}
}
\def\figiii{
\begin{figure}[t!] 
\centerline{
\epsfxsize=0.98\hsize
\epsfbox{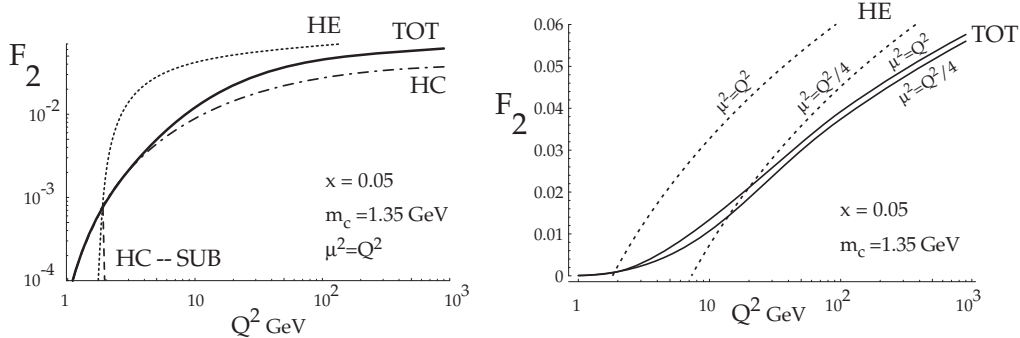}
}
\vspace{-00pt}
\caption{
 a) Same as Fig.~\protect\ref{fig2}, but with $\mu^2 = Q^2$. 
 b) $\mu$-variation in the  general mass formalism with mass-dependent 
evolution.  The variation of $F_2^{TOT}$ is small in contrast with that 
of $F_2^{HE}$.  Note,  we use a linear scale for $F_2$ to give a
proper perspective.
}
\vspace{-00pt}
\label{fig3}
\end{figure}
}

\title{
\vspace*{-1in}
\hfill\parbox[t]{1.3in}
{\normalsize
                                                       CTEQ-708        \\
                                                       SMU-HEP-9709    \\
                                                       hep-ph/9707459
}\\~\\~\\~\\ \bf   
  Heavy Quark Parton Distributions: 
{\Large 
   Mass-Dependent {\it or} Mass-Independent Evolution?}\footnote{%
This presentation is based on work done in collaboration with 
J.~C.~Collins and \hbox{Wu-Ki~Tung}. 
 This work is  supported by the U.S. Department of Energy
and the Lightner-Sams Foundation.
}
}

\author{ 
   Fredrick I. Olness and Randall J. Scalise 
}

\address{
   Department of Physics,
   Southern Methodist University, \\
   Dallas, Texas  75275-0175  USA
}

\maketitle

\begin{abstract} 
 In a consistently formulated pQCD framework incorporating non-zero
mass heavy quark partons, there is still the freedom to define
parton distributions obeying either  mass-independent or
mass-dependent evolution equations,  contrary to statements made in
a recent paper by MRRS.
 With properly matched hard cross-sections, different choices merely
correspond to different factorization schemes, and they yield the
same physical cross-sections. 
 We demonstrate this principle in a concrete order $\alpha_s$
calculation of the DIS charm structure function.
 We also examine the  proper matching between parton definitions and
subtractions in the hard cross-section near threshold where the
calculation is particularly  sensitive to mass effects of the heavy
quark. 
 The results obtained from the general-mass formalism are quite
stable against different choices of scale and  exhibit a smooth
transition in the threshold region  (using either mass-independent 
or mass-dependent evolution),  in contrast to results of
another recently proposed scheme.

\end{abstract}

 Recent improved measurements of heavy quark production
in both leptoproduction and hadroproduction demand better
theoretical understanding of the underlying QCD physics.
One must consider the changing role of the heavy
quark (denoted by $H$) from  threshold to the very high energy limit
available in current and future experiments. 
 Specifically, we address the question of the mass
dependence in the evolution of the parton distribution functions
(PDF's), $f_{a/P}(x,\mu )$.

 In the ACOT factorization scheme,\cite{acot,tung}
 the heavy quark mass ($m_H$) is fully incorporated in the hard
cross-section and threshold ($\mu=m_H$) 
matching conditions on $\alpha_s(\mu)$ and
$f_{a/P}(x,\mu)$.
  This scheme has the desirable feature that 
 the evolution of  all partons, including the heavy quark, 
is controlled by
the mass-independent \MSbar\  kernels; {\it i.e.},  
$f_{a/P}(x,\mu)$ satisfy the well established evolution equations. 
 This is a precise consequence of
the renormalization scheme in the theory with mass, {\it it is not
an approximation}.\footnote{%
 As the \MSbar\ scheme yields mass-independent renormalization
constants to all orders, the evolution kernels must be mass-independent.
 As $m_H \to 0$,   the massive hard scattering
$\sigma$'s  reduce to precisely the zero-mass \MSbar\ $\sigma$'s without
any finite renormalization;
 in this sense, the ACOT scheme is uniquely
the minimal massive extension of the \MSbar\ scheme.\cite{acot}
 }

A  general mass formalism  can, in principle, be implemented
with factorization prescriptions different from that of ACOT.
In a consistently formulated theory, physical predictions 
from different schemes
must remain the same  up to higher-order corrections.
 In a recent paper, MRRS \cite{MRRS} proposed a procedure
within the general mass approach involving {\it  mass-dependent splitting
functions} for the evolution equations. 
 They questioned the validity of the ACOT scheme because of its 
use of mass-independent evolution. 
 It is important to clarify this question and determine what schemes are
truly viable,  since further
development of heavy quark theory and phenomenology rely on the
availability of dependable calculations.

\section{Scheme Independence of Physical Cross Section}

We consider charm production in DIS 
in a  general renormalization/factorization scheme $R$.
The dominant contributions  
 are:\footnote{%
 We refer to the term in Eq.~\ref{eq:tung} as:
$\sigma_{TOT} = \sigma_{HE} + (\sigma_{HC} - \sigma_{SUB})$,
where $\sigma_{TOT}$ represents the total physical cross section,
$\sigma_{HE}$ the heavy-flavor excitation term ($\gamma^* c \to
c$),
$\sigma_{HC}$ the heavy-flavor creation term 
($\gamma^* g \to c \bar{c}$), and
$\sigma_{SUB}$ the subtraction term.
 }
\begin{eqnarray}
\sigma_{phys} &=&
{}^{R}f_{c/P} \otimes {}^{R}\widehat\sigma^{(0)}_{c\gamma^* \to c} 
+ 
{}^{R}f_{g/P} \otimes \phantom{(} {}^{R}\widehat\sigma^{(1)}_{g\gamma^* 
\to c\bar{c}} 
\phantom{-{}^{R}\widetilde\sigma^{1}_{g\gamma^* \to c\bar{c}} )\  }
+ 
{\cal O}(\alpha_s^2)
\nonumber \\
&=&
{}^{R}f_{c/P} \otimes \phantom{{}^{R}}\sigma^{(0)}_{c\gamma^* \to c} 
+ 
{}^{R}f_{g/P} \otimes ( 
\phantom{{}^{R}} \sigma^{(1)}_{g\gamma^* \to c\bar{c}}  
-
{}^{R}\widetilde\sigma^{(1)}_{g\gamma^* \to c\bar{c}}  
) 
+ 
{\cal O}(\alpha_s^2)
\label{eq:tung}
\end{eqnarray}
  ${}^{R}\widehat\sigma^n$ denotes the {\it hard cross section} in a 
given renormalization scheme $R$ at order $n$.
 $\sigma^n$ denotes the scheme independent 
 {\it partonic cross section}.\footnote{%
 To focus on the heavy quark parton singularities (of the form 
$\ln(m_H/Q)$), throughout the paper 
it is understood that
 the collinear singularities associated with light
partons are always subtracted from $\sigma^n$ 
according to the usual \MSbar\ scheme.
 }
 ${}^{R}\widetilde\sigma^n$ denotes the {\it subtraction} terms which 
removes the
potential heavy quark collinear singularities from $\sigma^n$.
 The physical cross section $\sigma_{phys}$ is scheme independent.
The choice of subtraction ${}^{R}\widetilde\sigma^n$  defines the scheme,
and hence the parton distributions ${}^{R}f_{a/P}$ and their associated 
splitting functions ${}^{R}P_{a\to b}$.
 In fact, to this order these terms are connected by
the relation: 
 $ {}^{R}\widetilde\sigma^{(1)}_{g\gamma^* \to c\bar{c}}  
 = 
{}^{R}f_{c/g}^{(1)} \otimes \sigma^{(0)}_{c\gamma^* \to c}
 $,
 where ${}^{R}f_{c/g}^{(1)}$ is the ${\cal O}(\alpha_s^1)$ perturbative
distribution of $c$ in $g$.
For the complete $\sigma_{SUB}$ we have:
 \begin{eqnarray}
\sigma_{SUB} &=& 
{}^{R}f_{g/P} \otimes 
{}^{R}\widetilde\sigma^{(1)}_{g\gamma^* \to c\bar{c}}  
= 
{}^{R}f_{g/P} \otimes
\frac{\alpha_s}{2 \pi}  
\int_{m_H^2}^{\mu^2}  \frac{d\mu^2}{\mu^2}  \ 
{}^{R}P^{(1)}_{g\to c} 
\otimes 
 \sigma^{(0)}_{c\gamma^* \to c}
\label{eq:sub}
 \end{eqnarray}
 Near threshold ($\mu \sim m_H$), we have
 \begin{eqnarray}
\sigma_{HE} &=& 
{}^{R}f_{c/P} \otimes {}^{R}\widehat\sigma^{(0)}_{c\gamma^* \to c} 
\simeq
{}^{R}f_{g/P} \otimes
\frac{\alpha_s}{2 \pi}  
\int_{m_H^2}^{\mu^2}  \frac{d\mu^2}{\mu^2}  \ 
{}^{R}P^{(1)}_{g\to c} 
\otimes 
 \sigma^{(0)}_{c\gamma^* \to c}
+ {\cal O}(\alpha_s^2)
\label{eq:he}
 \end{eqnarray}
which reflects the fact that 
$\sigma_{SUB} \simeq \sigma_{HE}$, by construction. 
 This  yields 
 $\sigma_{TOT} \simeq \sigma_{HC} + {\cal O}(\alpha_s^2)$ 
independent of the specific choice of scheme 
$R$, and therefore independent of the choice of 
 ${}^{R}\widetilde\sigma^{(1)}_{g\gamma^* \to c\bar{c}}$
 and ${}^{R}P^{(1)}_{g\to c}$.
 This is the key mechanism that compensates the different effects of
the mass-independent {\it vs.} mass-dependent evolution, and yields a
 $\sigma_{TOT}$ which is identical up to higher-order terms. 

\section{Mass-Dependent or Mass-Independent Evolution?}

\figi 

Ref.\cite{MRRS} strongly advocated the use of mass-dependent evolution of
the parton distributions, and questioned the correctness of using the
familiar \MSbar\ mass-independent evolution in ref.~\cite{acot}. 
Is this criticism justified? No--provided the proper matching between the 
hard scattering cross sections and the 
parton distributions is observed. Although the ACOT
scheme chose to use a {\it mass-independent} evolution for its simplicity
and natural connection to the familiar zero-mass parton results,
 this choice is not an approximation---the  physical predictions are
insensitive to this choice.
 We explicitly demonstrate this by implementing a mass-dependent evolution
scheme in the same formalism, and compare the new results with the existing
mass-independent ACOT calculation. 
 In the next section, we shall compare this new massive calculation with
that of MRRS.

In  Fig.~1  we display
the separate contributions to $F_2$  for both mass-independent 
and mass-dependent evolution. 
 The matching  properties discussed earlier are best examined
by comparing the (scheme-dependent) $F_2^{HE}$ and $F_2^{SUB}$ contributions
of  Fig.1a.
 We observe the following.
 {\it i)}~Within each scheme, $F_2^{HE}$ and $F_2^{SUB}$  are well
matched near threshold. 
 {\it ii)}~The matching of $F_2^{HE}$ and $F_2^{SUB}$ ensures that
the scheme dependence of $F_2^{TOT}$ is properly of higher-order
in $\alpha_s$.

The lesson is clear:  the choice  of a mass-independent
\MSbar\ or a mass-dependent (non-\MSbar) evolution is  purely a choice
of scheme, and   becomes simply a matter of
convenience--{\it there is no physically new information gained
from the mass-dependent evolution.}

\section{$F_2^{TOT}$ should match fixed-order result near threshold}
\nobreak

\figii 

A necessary consistency check  of any proposed general mass scheme in the 
important
intermediate energy region is that it yield the correct limits 
for threshold and asymptotic energies.
 Although this feature is, in principle, built into the formalism discussed
in Eq.~\ref{eq:sub},
 it is essential to verify that the expected behavior
actually appears in specific implementation of the scheme---in
particular, near the threshold region where the required cancellation
between the $F_2^{HE}$ and $F_2^{SUB}$ terms is delicate

 Fig.2 compares $F_2$ {\it vs.} the physical variable $Q$ using
the same charm mass, scale choice, and format of an existing MRRS
plot, ({\it cf.} Fig.6b, Ref.\cite{MRRS}).
 In Fig.2a, $F_2^{TOT}$ smoothly
interpolates between  $F_2^{HC}$ in the threshold region, and 
 $F_2^{HE}$ in the high energy region, demonstrating the
desired matching.\footnote{%
 The intuitive reason for these physical limits is identical  to
the  original mass-independent scheme.\cite{acot}\ 
 In the threshold region  the ``heavy quark" behaves more like an
extrinsic heavy object, while for high energies the ``heavy quark"
behaves like  an intrinsic parton.
 }
 In contrast, the corresponding curve in Fig.2b (taken from
Ref.\cite{MRRS}) shows a sudden rise at threshold  which indicates a
lack of expected cancellation between  $F_2^{HE}$ and $F_2^{SUB}$ 
just above threshold. 
 Since both calculations use the same (MRRS) evolution kernel, 
this difference  can only be due
to the implementation of the subtraction procedure: Fig.2a uses the
general mass formalism described above,\cite{acot,tung}; Fig.2b uses
the  procedure described in \hbox{Ref.\cite{MRRS}.}

\section{$F_2^{TOT}$ should be stable under scheme/scale choice}

\figiii 

Irrespective of smoothness of the theory calculation, a
physical prediction must be stable against the choice of
artificial theoretical parameters such as the renormalization
and factorization scale $\mu$. Fig.3a shows our calculation
(again, using the mass-dependent evolution) with a different choice
of $\mu$ compared to Fig.2a.
 In Fig.3b we note that while $F_2^{HE}$ changed dramatically with
choice of $\mu$,  this change is compensated by 
$F_2^{SUB}$  such that $F_2^{TOT}$ remains stable.

The corresponding comparison is not available for the MRRS
procedure. If the threshold behavior evident in Fig.2b
persists for other choices of scale, the position of the
sudden rise in cross-section will move with the choice of
scale; that would be cause for concern.\footnote{%
 For a scale choice of $\mu^2 = Q^2/k^2$, the charm quark evolution
turns on at $Q^2= k^2 m_c^2$, where $k^2$ is a arbitrary parameter. 
 }

\section{Summary}

In conclusion, we find that the presence or absence of heavy quark
mass in the evolution kernels has no physical effect  on
cross sections---there is
no physically new information gained from a mass-dependent evolution.

A mass-dependent evolution scheme has inherent difficulties. 
  The definition of the mass-dependent splitting kernels introduces a
kinematic ambiguity since it is impossible to
have an on-shell massless particle   split into two collinear
massive   on-shell  particles (e.g., $g \to c \bar{c}$); this
leads to a number of different definitions for the mass-dependent
splitting kernels.\cite{kernels}
 Since the PDF's in the mass-dependent scheme are not \MSbar, the hard
scattering cross sections $\hat{\sigma}$ must be converted from the
\MSbar\  scheme to this scheme. 
 At NLO, this conversion, as well as the PDF evolution, 
requires two-loop splitting kernels ($P^{(2)}$) computed in the
mass-dependent scheme; these results do not yet exist.

 The hallmark of any consistently formulated pQCD theory is: 
 {\it i)} proper matching between $F_2^{HE}$ and $F_2^{SUB}$ at
threshold, and 
 {\it ii)} insensitivity of the physical $F_2^{TOT}$   to the  arbitrary
scheme/scale choice. 
  The properly implemented general mass formalism  satisfies these criteria.
 The ACOT scheme (with mass-independent \MSbar\ evolution) has the added
advantage that the known mass-independent \MSbar\  kernels follow naturally
from the renormalization scheme adopted.

\null
\null

\end{document}